\documentclass[onecolumn,amsmath,amssymb,12pt,superscriptaddress,nofootinbib]{revtex4}
\pdfoutput=1

\usepackage[latin1]{inputenc}
\usepackage[english]{babel}
\usepackage{amssymb}
\usepackage{amsmath}
\usepackage{amsthm}
\usepackage[]{graphicx}
\usepackage[]{subfigure}
\usepackage{tensor}
\usepackage{color}
\usepackage{cancel}
\usepackage{setspace}
\usepackage{fancyhdr}
\usepackage{framed}
\usepackage{braket}
\usepackage{tikz}
\newcommand{\be}{\begin{equation}}
\newcommand{\ee}{\end{equation}}
\usepackage[bookmarks,linktocpage, colorlinks=true, plainpages = false, citecolor = magenta,  linkcolor=blue, urlcolor = magenta, filecolor = blue]{hyperref} 

\usepackage{mathrsfs}
\usepackage[title]{appendix}
\numberwithin{equation}{section}

\begin{document}

\allowdisplaybreaks

\title{Landau levels in a gravitational field: The Levi-Civita and Kerr spacetimes case}
\vspace{.3in}

\author{Fay\c{c}al Hammad}
\email{fhammad@ubishops.ca}
\affiliation{Department of Physics and Astronomy, Bishop's University, 2600 College Street, Sherbrooke, QC, J1M~1Z7
Canada}
\affiliation{Physics Department, Champlain 
College-Lennoxville, 2580 College Street, Sherbrooke,  
QC, J1M~0C8 Canada}
\affiliation{D\'epartement de Physique, Universit\'e de Montr\'eal,\\
2900 Boulevard \'Edouard-Montpetit,
Montr\'eal, QC, H3T 1J4
Canada} 

\author{Alexandre Landry} \email{alexandre.landry.1@umontreal.ca} 
\affiliation{D\'epartement de Physique, Universit\'e de Montr\'eal,\\
2900 Boulevard \'Edouard-Montpetit,
Montr\'eal, QC, H3T 1J4
Canada}

\begin{abstract}
We have recently found that the gravitational field of a static spherical mass removes the Landau degeneracy of the energy levels of a particle moving around the mass inside a magnetic field by splitting the energy of the Landau orbitals. In this paper we present the second part of our investigation of the effect of gravity on Landau levels. We examine the effect of the gravitational fields created by an infinitely long massive cylinder and a rotating spherical mass. In both cases, we show that the degeneracy is again removed thanks to the splitting of the particle's orbitals. The first case would constitute an experimental test --- which is quantum mechanical in nature --- of the gravitational field of a cylinder. The approach relies on the Newtonian approximation of the gravitational potential created by a cylinder but, in view of self-consistency and for future higher-order approximations, the formalism is based on the full Levi-Civita metric. The second case opens up the possibility for a novel quantum mechanical test of the well-known rotational frame-dragging effect of general relativity.\\
\end{abstract}

\maketitle
\tableofcontents


\section{Introduction}\label{sec:I}
We have recently shown in Ref.~\cite{GravityLandau} that gravity has a nontrivial effect on the quantum Landau energy levels (see, e.g., Ref.~\cite{LandauLifshitz}) of a particle moving around a spherical static mass and surrounded by a constant and uniform magnetic field. We found that the degeneracy of the Landau levels is removed by splitting the energy of each of the Landau orbitals. We have also pointed out that the gravitational splitting of the levels could be used to test departures from the inverse square-law of gravity using quantum particles.

It is worth recalling, and emphasizing, here that the investigation done in Ref.~\cite{GravityLandau} belongs actually to two different classes of research involving gravity and quantum particles. The first class of investigations aims at bringing into light the gravitational effects on a quantum particle. Noticeable among such investigations are those studying the behavior of cold neutrons inside a gravitational field \cite{UCNinEarth1,UCNinEarth2,UCNinEarth3,BookNeutronsInterfro,Kulin,Abele,Landry1,Landry2}. The central idea behind such an approach is to substitute the gravitational potential for the usual electric potentials frequently used in non-relativistic quantum mechanics of point particles. Just like with the familiar electric potentials, the gravitational one is indeed expected to induce a quantization of the energy of the particle immersed inside it, or even lead to interference patterns as the particle moves inside such a potential. From this point of view, the work we presented in Ref.~\cite{GravityLandau} indeed subscribes to this class of research by showing that the effect of the additional potential due to gravity modifies the familiar Landau energy levels of a charged particle inside a magnetic field.

Besides showing how gravity disturbs the Landau levels of a charged particle, however, the work in Ref.~\cite{GravityLandau} allowed to explicitly expose the effect a departure from the inverse square-law for gravity would have on a quantum particle. As such, our previous work can also be assigned to the second class of investigations the aim of which is to {\it test} the gravitational field itself in the same spirit as the works such as the one in Ref.~\cite{Biedermann}. Investigating the effect of gravity on Landau levels becomes thus a fundamental approach in the sense that it consists in using the {\it quantum} theory to probe the {\it classical} gravitational field.

In the present paper, our goal is to explore even more this second category of investigations while still providing evidence for the effect of more complicated spacetimes on a quantum particle. More specifically, we are going to study the fate of Landau's energy levels of a charged particle inside a uniform and constant magnetic field, first (i) when the particle is moving within the Levi-Civita spacetime then (ii) when the particle is moving within the Kerr spacetime. The goal of the first investigation is, above all, to contribute to the existing efforts in the literature towards devising tests for that elusive and much debated spacetime of general relativity. In fact, although the so-called Levi-Civita spacetime was discovered exactly now a century ago \cite{Weyl,Levi-CivitaPaper}, such a metric still holds many mysteries and is less often used in the literature compared to the more famous ones of the Schwarzschild and Kerr spacetimes.

The Kerr metric, in contrast, is indeed very well known, for it is used mainly to describe the spacetime of a rotating black hole. As such, our present investigation based on the Kerr metric does actually more than just test a special solution to the gravitational equations. It provides a novel way --- {\it quantum mechanical} in nature --- for testing the frame-dragging effect of general relativity which, hitherto, has only been tested through the famous Lense-Thirring effect (see e.g., Ref.~\cite{Iorio} and the references therein). The Lense-Thirring effect consists of the precession of a gyroscope, or any spinning body, in the vicinity of a rotating mass, like the Earth. In this paper, we show how frame-dragging creates a specific signature on the splitting of the quantum Landau levels. For approaches relying instead on the effect of rotating frames on the quantum spin of particles and on their internal clocks, as well as on quantum interferences, see Refs.~\cite{Dowker,Mashhoon1,Lammerzahl,Herrera,Tartaglia,Mashhoon2,Ruggiero,Okawara1,Okawara2,Okawara3,Interference}.    

The remaining sections of this paper are organized as follows. In Section~\ref{sec:II}, we build the curved-spacetime Klein-Gordon equation for a charged particle minimally coupled to the electromagnetic field and moving in the full Levi-Civita spacetime created by an infinitely long massive cylinder. We then solve the equation in the Newtonian approximation, with the goal of making contact with laboratory experimental tests, and then we evaluate the splitting of the Landau levels. In Section~\ref{sec:III}, we repeat the same analysis as the one done for the Levi-Civita spacetime in Section~\ref{sec:II}, but using the Kerr metric instead. We compare the Landau levels splitting caused by the latter spacetime to the one obtained in Ref.~\cite{GravityLandau} within the static spherical mass. The frame-dragging effect reveals itself clearly. We conclude this paper with a brief discussion and conclusion section. A short appendix is included in which many of the complicated integrals needed in the text are gathered for reference.
\section{A particle inside a magnetic field in the Levi-Civita spacetime}\label{sec:II}

In both this section and the next, we are going to use a charged spinless particle of mass $m$ as our test particle. For practical purposes, we are going to assume the particle has the elementary charge $e$. This is motivated by the possibility of experimentally implementing the setup by using heavy ions the spin of which can be neglected. Thus, the Klein-Gordon equation for a scalar field in curved spacetime, $\left(\Box+m^2c^2\right)\varphi=0$, will be amply sufficient for our present purposes. 

For a minimally coupled particle to the electromagnetic field $A_\mu$ and to the metric $g_{\mu\nu}$ of the spacetime, the Klein-Gordon equation reads (see, e.g., Ref.~\cite{Kiefer}),
\begin{equation}\label{KG}
\left[\frac{1}{\sqrt{-g}}D_\mu\left(\sqrt{-g}g^{\mu\nu}D_\nu\right)+m^2c^2\right]\varphi(x)=0,    
\end{equation}
where, the minimal-coupling prescription, $D_\mu=-i\hbar\partial_\mu-eA_\mu$, is assumed. Next, the Levi-Civita metric around an infinitely long cylinder of mass $M$ can be written in the cylindrical coordinates $(t,\rho,\phi,z)$ as follows \cite{Fulling}:
\begin{equation}
\label{LCMetric}
{\rm d}s^2=-c^2\left(\frac{\rho}{\rho_*}\right)^{-2a}{\rm d}t^2+\left(\frac{\rho}{\rho_*}\right)^{-2(a+b)}{\rm d}\rho^2+K^2\rho^2{\rm d}\phi^2+\left(\frac{\rho}{\rho_*}\right)^{-2b}{\rm d}z^2.    
\end{equation}
Here, the constants $a,b$ and $K$ are all arbitrary --- with a constraint between $a$ and $b$ --- while $c$ is the speed of light. Note that the form (\ref{LCMetric}) of the Levi-Civita metric we use here is not the one that one often encounters in the literature \cite{Marder,Bonnor,Herrera2,Santos} (see also, the very nice recent review \cite{Bronnikov}.)
In fact, the first difference is that we have introduced here the fixed constant radius $\rho_*$ to allow us to keep inside the metric the radius $\rho$, describing the position of the particle away from the center of the cylinder, with the dimensions of a length. In addition, in view of the approximations we are going to perform in order to be able to solve our differential equations, having a dimensionless ratio is very well suited for expanding the metric in powers of such a ratio as well as for keeping the argument of the logarithms appearing there dimensionless.
Furthermore, as is well-known in the case of logarithmic potentials, in particular the one created by an infinitely long cylinder (see, e.g., Ref.~\cite{BookPotentialCylinder}), one does not have a vanishing potential anywhere. As a consequence, the reference point for potentials cannot be taken to be at infinity anymore (as is the case with a spherical mass). To remedy such an issue, one introduces a fixed radial distance from the center of the source and takes such a point to be the reference for measuring potentials. As we shall see shortly, our fixed radius $\rho_*$ allows us specifically to take it to be the reference point of zero gravitational potential. 

The second difference with respect to the usual forms of the metric given in the literature is the presence of the multiplicative constant $K$ in the angular component of the metric. The role of this constant is actually just to avoid rescaling the azimuth angle $\phi$, keeping it instead within the familiar range $[0,2\pi[$ \cite{Fulling}. As such, any possible excess or deficit angle, which would give rise to a conical singularity whenever the spacetime around the cylinder is continued all the way to the center of the latter (the latter becoming then a string), is encoded in the constant $K$. If $K<1$, one has a deficit angle (a wedge is removed from spacetime), whereas for $K>1$ one has an excess of spacetime (a wedge is added)\footnote{The role of this parameter in the Levi-Civita metric in encoding the global topology of the spacetime was first pointed out in Ref.~\cite{Bonnor}.}. As we shall see below, our results make even more transparent the effect of this deficit/excess angle on the energy-spectrum of the particle when we display explicitly the constant $K$ in the metric instead of absorbing it by redefining the angle $\phi$. 

The third difference in our choice (\ref{LCMetric}) for the form of the Levi-Civita metric, is the fact that our $z$- and $\rho$-component of the metric acquire different coefficients. This specific choice is merely made here for the sake of simplicity. In fact, had we chosen to use instead the more familiar form of the metric, in which both coordinates acquire the same metric component \cite{Santos,Bronnikov}, the angular component of the metric would also acquire\footnote{See Ref.~\cite{Fulling} for the various coordinate re-definitions that allow one to switch from one form of the metric to the other.} a power function of $\rho$ instead of having the above familiar factor $\rho^2$. This would indeed only render our equations and analysis uselessly complicated. 

Let us now focus on the meaning of the remaining two parameters $a$ and $b$ of the metric. First, as mentioned above, the two parameters are not completely arbitrary as they obey a specific constraint in the form of an algebraic relation between them. Such a relation reads, $ab+a+b=0$ \cite{Fulling}. This implies that the total number of independent parameters in the Levi-Civita spacetime is actually just two. The meaning of one of these two parameters, say, $a$, can now be found as follows. For very small $a$, we can expand the $00$-component of the metric to the first order as, $(\rho/\rho_*)^{-2a}\approx 1-2a\ln(\rho/\rho_*)$. Comparing this with the weak-field approximation of general relativity, $g_{00}\approx-1+2U$, reveals what potential $U$ in the post-Newtonian approximation one has; it reads $U=a\ln(\rho/\rho_*)$. This, when compared, in turn, with the well-known Newtonian potential around an infinitely long cylinder \cite{BookPotentialCylinder}, implies that $a$ should be identified with the product $G\lambda/c^2$, where $\lambda$ is the linear mass density of the cylinder. For a finite-radius cylinder, we should then identify $a$ with $\pi G\gamma\rho_0^2/c^2$ when the radius of the cylinder is $\rho_0$ and its volume mass density is $\gamma$. It should be kept in mind, though, that for the infinitely-long cylinder approximation to be accurate in the case of a finite cylinder, the particle should be kept very close to the lateral surface of the long cylinder.

It is worth noting here, however, that, as alluded to in the Introduction, the Levi-Civita metric is not free from ambiguities when it comes to its full interpretation. In fact, it was shown in Refs.~\cite{Herrera2,Santos} that only for the range $0<a<1/4$ of the parameter $a$ does one extract a physically sensible spacetime around the cylinder, for only then do circular time-like geodesics exist. For $a=1/4$ or $a=1$, the circular geodesics are null, whereas for $1/4<a<1$ those geodesics are spacelike. The circular geodesics become timelike only for $0<a<1/4$ or $a>1$. Fortunately, since we are interested here in the case $a\ll1$, such issues do not need to worry us.
Nevertheless, these serious obstacles in the interpretation of the Levi-Civita metric make actually the investigation of the effects of the metric on quantum particles, not only a way for testing cylindrical gravitational fields, but constitutes thus an additional input towards understanding the metric itself.

Let us now substitute the metric (\ref{LCMetric}) into the Klein-Gordon equation (\ref{KG}). As for the vector potential $A_\mu$, we are going to use the usual symmetric gauge adapted to the cylindrical coordinates $(t,\rho,\phi,z)$ in which the only non-vanishing component reads\footnote{We display here the covariant form of the potential vector, as the tetrad form we used in the first version of this manuscript leads to much confusion. In fact, while the usual tetrad form of the vector potential $A_{\hat{\phi}}=\frac{1}{2}B\rho$ has the advantage of displaying the right dimensions for a potential vector, it requires one to be careful when substituting it inside Eq.~(\ref{KG}). By taking such care, the result one finds is, of course, the same with both expressions.}, $A_\phi=\frac{1}{2}KB\rho^2$. Note that with the presence of the magnetic field $\bf B$, one might expect a spacetime metric that is not the one in Eq.~(\ref{LCMetric}), but a metric that would be a solution to the Einstein-Maxwell equations with a massive infinitely long cylinder. However, as explained in detail in Ref.~\cite{GravityLandau}, we assume the magnetic field to be as weak as to allow us to neglect its {\it geometric} effect on the spacetime and, hence, neglect its effect on the particle via geometry. In fact, the correction that arises from taking into account the effect of the magnetic field on geometry is of the order of $G\epsilon_0c^{-2}B^2\rho^2$ and becomes significant only for magnetic fields of the order of $10^{19}{\rm G}$ \cite{DiracInMelvin}. We therefore focus in this paper only on the effect of the magnetic field on the particle due the usual Lorentz force.

Now, because of the time-independence of both the metric and the magnetic field, and thanks to the symmetry of the planar motion of the particle around the $z$-axis, we expect the wavefunction for the particle of energy $E$ to be of the form,  $\varphi(t,\rho,\phi,z)=e^{-i\frac{Et}{\hbar}}e^{i\ell\phi}R(\rho)$, with $\ell$ a non-negative integer. For simplicity, we assume that the particle has no momentum along the $z$-direction and that it is moving counterclockwise around the cylinder. Therefore, the Klein-Gordon equation (\ref{KG}) in the curved spacetime (\ref{LCMetric}) takes the form,
\begin{equation}\label{KGinLC1}
\Bigg[\frac{E^2}{\hbar^2c^2}\left(\frac{\rho}{\rho_*}\right)^{2a}-\frac{m^2c^2}{\hbar^2}+\left(\frac{\rho}{\rho_*}\right)^{2a+2b-1}\partial_\rho\left(\frac{\rho}{\rho_*}\,\partial_\rho\right)-\frac{\ell^2}{K^2\rho^2}+\frac{eB\ell}{K\hbar}-\frac{e^2B^2\rho^2}{4\hbar^2}\Bigg]R(\rho)=0.
\end{equation} 
In the case of small parameters, $a,b\ll1$, the powers of the ratio $(\rho/\rho_*)$ can be expanded and the above equation then reads, at the first-order approximation in $a$ and $b$, as follows:
\begin{multline}\label{KGinLC2}
\frac{{\rm d}^2R}{{\rm d}\rho^2}+\frac{{\rm d}R}{\rho{\rm d}\rho}+\Bigg[\left(\frac{E^2}{\hbar^2c^2}-\frac{m^2c^2}{\hbar^2}-\frac{\ell^2}{K^2\rho^2}+\frac{eB\ell}{K\hbar}-\frac{e^2B^2\rho^2}{4\hbar^2}\right)\\
+2\left(\frac{m^2c^2}{\hbar^2}+\frac{\ell^2}{K^2\rho^2}-\frac{eB\ell}{K\hbar}+\frac{e^2B^2\rho^2}{4\hbar^2}\right)(a+b)\ln\left(\frac{\rho}{\rho_*}\right)-\frac{2bE^2}{\hbar^2c^2}\ln\left(\frac{\rho}{\rho_*}\right)\Bigg]\,R(\rho)=0.
\end{multline}
Next, performing the change of variable $R(\rho)=\psi(\rho)/\sqrt{\rho}$, and then decomposing the total energy of the test particle as, $E=\mathcal{E}+mc^2$, and using the non-relativistic approximation $E^2\approx2mc^2\mathcal{E}+m^2c^4$, the above equation, in turn, simplifies to,
\begin{equation}\label{KGinLC3}
-\frac{\hbar^2}{2m}\psi''+\Bigg[\frac{e^2B^2\rho^2}{8m}+\frac{\hbar^2}{2m\rho^2}\left(\frac{\ell^2}{K^2}-\frac{1}{4}\right)-\frac{\hbar eB\ell}{2mK}-amc^2\ln\left(\frac{\rho}{\rho_*}\right)\Bigg]\psi=\mathcal{E}\psi.
\end{equation}
We have denoted by a prime a derivative of $\psi(\rho)$ with respect to the radial variable $\rho$. In addition, we have kept here only the leading term $mc^2$ from the second and third lines of Eq.~(\ref{KGinLC2}). Again, this approximation is amply sufficient for our purposes here, for we have indeed $\hbar eB/m\ll mc^2$ for the orders of magnitude of the magnetic fields used in the laboratory. This Schr\"odinger equation will give us the full quantized energy spectrum of the particle.

Now, we argued at length in Ref.~\cite{GravityLandau} (see also the references therein) that there are essentially two practical working methods for extracting the quantized energy levels from such a Schr\"odinger equation containing extra non-trivial central potentials. Our non-trivial extra term here is the logarithmic term inside the square brackets of Eq.~(\ref{KGinLC3}). The first approach relies on the time-independent perturbation theory. The second approach consists simply in approximating the effective potential, contained inside the square brackets, by that of a simple harmonic oscillator. When using the latter approach, one would directly read off the energy levels as given by the familiar energy spectrum of a simple harmonic oscillator. We are going to apply in the rest of this section both methods, starting with the one relying on the time-independent perturbation theory. A short note will be given at the very end of this section about two other less reliable and less practical methods for extracting the quantized levels.
\subsection{Using Perturbation theory}
In Ref.~\cite{GravityLandau}, we have already found the solutions to Eq.~(\ref{KGinLC3}) without the very last term inside the square brackets. Those solutions constitute the unperturbed eigenvalues of the Landau Hamiltonian. Note, however, that now we have the extra parameter $K$ that enters even in the unperturbed equation. Nevertheless, the solutions with such an extra parameter can easily be adapted from the results of Ref.~\cite{GravityLandau}. Indeed, this can be accomplished simply by introducing the reduced orbital quantum number, $\bar{\ell}=\ell/K$. For simplicity, however, and without any loss of generality, we are going to set hereafter $K=1$. The effect of the parameter $K$, when the latter is different from unity, can then be inferred from the results with $K=1$ just by replacing $\ell$ by $\bar\ell$.

The unperturbed eigenfunctions $\psi^{(0)}_{n\ell}(\rho)$ corresponding to Eq.~(\ref{KGinLC3}) without the last term inside the square brackets are then \cite{GravityLandau},
\begin{equation}\label{UnperturbedWF}
\psi^{(0)}_{n\ell}(\rho)=A_{n\ell}\;\rho^{\ell+\frac{1}{2}}e^{-\frac{\beta}{4}\rho^2}\,_1F_1\left(-n;\ell+1;\frac{\beta}{2}\rho^2\right).
\end{equation}
The special functions $\,_1F_1(a,b,z)$ are called Kummer's, or confluent hypergeoemtric, functions \cite{BookKummer}. As usual, $n$ is here a non-negative integer. The normalization constants $A_{n\ell}$ would, in principle, be determined by imposing as usual the completeness condition on the eigenfunctions, $\int_0^\infty \psi_{n\ell}^{(0)*}(\rho)\psi_{m\ell}^{(0)}(\rho) \,{\rm d}\rho=\delta_{nm}.$ However, in contrast to what is assumed in the case of cosmic strings, a cylinder of mass $M$ has a finite nonzero radius $\rho_0$. As a consequence, the test particle's position is necessarily limited to the interval of radii $\rho\in[\rho_0,\infty)$. In addition, our gravitational field is valid only for $\rho>\rho_0$, {\it i.e.}, outside the cylinder.

Because of this particular configuration, we should distinguish two different regions when solving the Schr\"odinger equation. The region outside the cylinder, for which $\rho>\rho_0$, and the region inside the cylinder, for which $\rho<\rho_0$. We shall assume, however, that the cylinder is completely reflective to the test particle. In other words, we take the particle's wavefunction to vanish inside the cylinder, meaning that the particle has zero probability of penetrating inside the latter. In fact, with this assumption we are simply dealing with a semi-infinite potential well, for then our system just consists effectively of a test particle moving around an infinitely long cylinder, inside of which the potential is infinite and outside of which the potential is gravitational and is given by Eq.~(\ref{KGinLC3}). The wavefunction outside the cylinder having the expression (\ref{UnperturbedWF}), all we need to further impose on the latter is its continuity across the surface $\rho=\rho_0$. This condition translates then into the requirement, $\psi^{(0)}_{n\ell}(\rho_0)=0$. Based on expression (\ref{UnperturbedWF}), this requirement is equivalent to the following identity to be imposed on the confluent hypergeometric function:
\begin{equation}\label{FContinuity}
    \,_1F_1\left(-n;\ell+1;\frac{\beta}{2}\rho_0^2\right)=0.
\end{equation}
This condition already arose for the case of a spherical mass examined in Ref.~\cite{GravityLandau}. Its physical interpretation is therefore similar to the one proposed in that reference. Indeed, this condition is due to the geometry of the system itself. The condition (\ref{FContinuity}) involves the two unknown integers $n$ and $\ell$ and, hence, implies that the latter are related to the parameter $\beta$, {\it i.e.}, the magnetic field, and to the radius $\rho_0$ of the cylinder. In the absence of the cylinder, all possible Landau levels $n$ and all possible orbital numbers $\ell$ would be accessible to the particle without any restriction. The presence of the cylinder at the center of motion disturbs the motion of the test particle by creating the forbidden region $0\le\rho\leq\rho_0$, implying that, depending on the value of the product $\tfrac{1}{2}\beta\rho_0^2$, a specific correlation emerges between the values of $n$ and $\ell$. This means that only specific combinations of the magnetic field and the radius of the cylinder, sitting at the center of motion of a test particle, would give rise to the quantum numbers $n$ and $\ell$ that the particle could take while moving around the cylinder and avoiding the interior of the latter.

In the case of a string-like mass distribution, {\it i.e.}, for $\rho_0=0$, the requirement $\psi^{(0)}_{n\ell}(\rho_0)=0$ is, of course, automatically satisfied. In that case, the condition (\ref{FContinuity}) does not need to be imposed anymore and, hence, no correlation between the quantum numbers $n$ and $\ell$ and the parameter $\tfrac{1}{2}\beta\rho_0^2$ is implied either. Since we are interested here only in the fate of the Landau energy levels inside the gravitational field, we are going to ignore in the remainder of this paper such a restriction and assume that a specific combination of the radius of the cylinder and of the magnetic field, guaranteeing the appearance of Landau quantum levels and orbitals for the particle, has already been set up.

Because of this forbidden region to the test particle, the normalization condition that we should imposed here is then $\int_{\rho_0}^\infty \psi_{n\ell}^{(0)*}(\rho)\psi_{n\ell}^{(0)}(\rho) \,{\rm d}\rho=1.$ 
In Ref.~\cite{GravityLandau}, the normalization constants of the wavefunctions $\psi_{n\ell}^{(0)}(\rho)$ implied by such a condition were explicitly found to be $A_{n\ell}=\mathcal{M}_{n\ell}^{-1/2}$. The quantities $\mathcal{M}_{n\ell}$ are infinite series obtained by setting $n=m$ in the infinite series $\mathcal{M}_{mn\ell}$, given explicitly for reference in Eq.~(\ref{AppendixIntegralM}) of appendix \ref{A}.
In addition, the energy eigenvalues corresponding to the unperturbed wavefunctions (\ref{UnperturbedWF}) are given by \cite{GravityLandau},
\begin{equation}\label{Landau}
\mathcal{E}_n^{(0)}=\frac{\hbar eB}{m}\left(n+\frac{1}{2}\right).
\end{equation}
These are the familiar Landau quantized energy levels. The high degeneracy of the levels shows up in the freedom the particle has with the orbital quantum number $\ell$ for each quantum number $n$. Note that, had we kept the parameter $K$ of the Levi-Civita metric (\ref{LCMetric}), these energy levels would not have been modified as the only difference would be the substitution $\ell\rightarrow\ell/K$.

The perturbed Landau energy levels due to the cylindrical gravitational field are now easy to compute at the first order using the time-independent perturbation theory. Although the Landau energy levels are infinitely degenerate, the fact that the gravitational interaction potential $V(\rho)=-amc^2\ln(\rho/\rho_*)$ around the cylinder is rotational symmetric means that the gravitational perturbation does not couple between two different Landau orbitals of quantum numbers $\ell$ and $\ell'$. This implies, as was the case with a spherical mass \cite{GravityLandau}, that the perturbation matrix elements  $\braket{n,\ell|V(\rho)|n,\ell'}$ are diagonal. Consequently, the degenerate time-independent perturbation theory yields the first-order correction, $\mathcal{E}_{n\ell}=\mathcal{E}_{n}^{(0)}+\braket{n,\ell|V(\rho)|n,\ell}$, where the term $\mathcal{E}_n^{(0)}$ represents the unperturbed $n^{\rm th}$ Landau level (\ref{Landau}). We have thus the following more explicit first-order correction to the energy of the $n^{\rm th}$ Landau level in the quantum orbital $\ell$: 
\begin{align}\label{LCellCorrection}
\mathcal{E}_{n\ell}&=\mathcal{E}_{n}^{(0)}-amc^2\int_{\rho_0}^{\infty} \psi^{(0)*}_{n\ell}(\rho)\psi^{(0)}_{n\ell}(\rho)\ln\left(\frac{\rho}{\rho_*}\right)\,{\rm d}\rho.
\end{align}
In order to evaluate the improper integral in this equation, we have to substitute expression (\ref{UnperturbedWF}) for the unperturbed wavefunctions $\psi_{n\ell}^{(0)}(\rho)$ and replace the normalization constants $A_{n\ell}$ by their expressions $\mathcal{M}_{n\ell}^{-1/2}$ as given by Eq.~(\ref{AppendixIntegralM}). Afterwards, by using the result (\ref{AppendixIntegralLDef}) of appendix \ref{A}, we find,
\begin{equation}\label{ExplicitellCorrection}
\mathcal{E}_{n\ell}=\mathcal{E}_{n}^{(0)}-amc^2\bar{\mathcal{L}}_{n\ell}\bar{\mathcal{M}}_{n\ell}^{-1}.
\end{equation}
As was done in Ref.~\cite{GravityLandau}, we have denoted here by $\bar{\mathcal{L}}_{n\ell}$ and $\bar{\mathcal{M}}_{n\ell}$ the reduced forms of the series (\ref{AppendixIntegralM}) and (\ref{AppendixIntegralL}), obtained by suppressing the constant factor $(2/\beta)^{\ell+1}$ common to both series, and by setting $n=m$ in both. This result shows how the degenerate Landau levels split at the first-order in $amc^2$ due to gravity. Although not explicitly displayed, the dependence of this splitting on the magnetic field $B$ is still present inside the individual series (\ref{AppendixIntegralM}) and (\ref{AppendixIntegralL})

For the sake of concreteness, let us evaluate the explicit correction to the first Landau level by setting $n=1$ in Eq.~(\ref{ExplicitellCorrection}). First, it is obvious from the defining integrals (\ref{AppendixIntegralMDef}) and (\ref{AppendixIntegralLDef}) of $\mathcal{M}_{n\ell}$ and $\mathcal{L}_{n\ell}$, respectively, that for small values of $\ell$, the gravitational correction to the first Landau level is of the order $-amc^2\ln(\rho_0/\rho_*)$. For larger values of $\ell$, however, one cannot easily get a simple physical picture of the effect of the gravitational field on the first Landau level based on the full expression (\ref{M1ell}) of $\mathcal{M}_{1\ell}$ and the full expression (\ref{L1ell}) from which $\mathcal{L}_{1\ell}$ can be found. For this reason, we are going instead to give here an estimate of the perturbation correction for the large-$\ell$ orbitals. In fact, in this case the expressions simplify greatly by using the asymptotic results (\ref{M1ellInfinite}) and (\ref{L1ellInfinite}) for $\mathcal{M}_{1\ell}$ and $\mathcal{L}_{1\ell}$, respectively. We find,
\begin{equation}\label{1ellCorrection}
\mathcal{E}_{1,\ell\gg1}=\mathcal{E}_{1}^{(0)}-amc^2\mathcal{L}_{1,\ell\gg1}\mathcal{M}_{1,\ell\gg1}^{-1}\approx\frac{3\hbar eB}{2m}+\frac{amc^2}{2}\ln\left(\frac{eB\rho_*^2}{2\hbar\ell}\right).
\end{equation}
This result shows that, just like what happens in the case of a spherical static mass \cite{GravityLandau}, the splitting brought to the Landau levels by the gravitational field of the cylinder has, in fact, a simple form for large orbitals $\ell$. This splitting is independent of the radius of the cylinder $\rho_0$ and depends instead on the fixed radius $\rho_*$ we took as a reference for the gravitational potential. In contrast to the case of the spherical mass \cite{GravityLandau}, however, the splitting depends here logarithmically on the magnetic field. For the case of $K\neq1$, the Landau term remains unaffected but the correction term does get affected as the denominator inside the logarithm acquires the multiplicative factor $K^{-1}$. On the other hand, as is the case with the spherical mass, from the general formula (\ref{ExplicitellCorrection}) we see that for large $n$, the first-order correction does not get suppressed.

It is worth noting here also that, like with the case of the spherical mass \cite{GravityLandau}, in the absence of the magnetic field, {\it i.e.} when setting $B=0$ in Eq.~(\ref{ExplicitellCorrection}), the first-order perturbation vanishes together with the zeroth-order levels $\mathcal{E}_{n}^{(0)}$, for both series $\mathcal{M}_{mn\ell}$ and $\mathcal{P}_{mn\ell}$ do not exist in this case as the integrals that gave rise to them vanish for $\beta=0$. A proper treatment of the motion of the particle around the cylinder without the magnetic field consists in solving the Schr\"odinger equation with only the logarithmic potential as the unique potential (see, e.g., Ref.~\cite{LogarithmicPotential}).

At the second order, the corrections to the energy levels would be even more complicated than what was found for the spherical mass case in Ref.~\cite{GravityLandau}. In fact, the correction $\mathcal{E}_{n\ell}^{(2)}=\sum\limits_{k\neq n}|\braket{k,\ell|V|n,\ell}|^2/(\mathcal{E}_{k}^{(0)}-\mathcal{E}_{n}^{(0)})$, which is quadratic in the product $am$, would involve, besides terms logarithmic in the magnetic field, a ratio with the magnetic field in the denominator as well. Suffice it then to note here that, like in the spherical mass case \cite{GravityLandau}, the second-order correction to the energy levels of the particle is quadratic in $G\lambda$, where $\lambda$ is the linear mass density of the long cylinder. Furthermore, because of the presence of the magnetic field in the denominator in such a correction, the latter is not valid anymore without the magnetic field, {\it i.e.}, when $B=0$. In this case, one should instead solve Eq.~(\ref{KGinLC3}) by setting $B=0$ there. In fact, in that case such an equation solves differently from the way the Schr\"odinger equation of the hydrogen atom is solved (see, e.g., Ref.~\cite{BookQM}). Different specific approximation methods can indeed then be applied for that case \cite{LogarithmicPotential}. We are not going here to deal with such a purely gravitational problem, for our main purpose in the present paper is the effect of gravity on the Landau levels.
\subsection{Using the harmonic oscillator approximation}
It is actually possible to also achieve quantization of the energy levels of the particle by starting from the Schr\"odinger equation (\ref{KGinLC3}) and approximating the latter with the equation of a harmonic oscillator. All one needs to do is find the equilibrium radius $\rho_{\rm e}$ around which the particle's effective potential $V_{\rm eff}(\rho)$, as given by the square brackets in Eq.~(\ref{KGinLC3}), reaches a minimum. Such a radius $\rho_{\rm e}$ is thus the solution to the equation ${\rm d}V_{\rm eff}(\rho)/{\rm d}\rho=0$. The latter equation is a quartic equation but its special form (quadratic in $\rho_e^2$),
\begin{equation}\label{VRho*}
\frac{e^2B^2}{4m}\rho^4_{\rm e}-amc^2\rho^2_{\rm e}-\frac{\hbar^2}{m}\left(\ell^2-\frac{1}{4}\right)=0,
\end{equation}
in contrast to the case of the effective potential around the spherical mass \cite{GravityLandau}, is easily solvable. It has indeed two roots; the positive one being,
\begin{equation}\label{RhoEq}
\rho_{\rm e}^2=\frac{2am^2c^2}{e^2B^2}+\sqrt{\frac{4a^2m^4c^4}{e^4B^4}+\frac{4\hbar^2}{e^2B^2}\left(\ell^2-\frac{1}{4}\right)}\approx\frac{2\hbar}{eB}\sqrt{\ell^2-\frac{1}{4}}\left(1+\frac{am^2c^2}{\hbar eB\sqrt{\ell^2-\frac{1}{4}}}\right).
\end{equation}
In the second step we have expanded in powers of $a$ up to the first order as this will allow us shortly to (i) easily see how one recovers the Minkowski case $a=0$, as well as to (ii) extract the first-order correction in $a$ to the Landau levels. The effective potential $V_{\rm eff}(\rho)$ of the particle around this equilibrium position $\rho_{\rm e}$ can now be Taylor-expanded at the second order in $\rho$ and approximated by a quadratic potential as follows:
\begin{equation}\label{VTaylorexpanded}
V_{\rm eff}(\rho)\simeq V_{0}+\frac{1}{2}m\omega^2(\rho-\rho_{\rm e})^2. 
\end{equation}
Here, $V_0=V_{\rm eff}(\rho_{\rm e})$ and $m\omega^2={\rm d}^2V_{\rm eff}/{\rm d\rho^2}|_{\rho=\rho_{\rm e}}$. With such a potential, the Schr\"odinger equation (\ref{KGinLC3}) becomes that of a simple harmonic oscillator for which the energy eigenvalues are well-known, and given by,
\begin{equation}\label{EnergySHO}
\mathcal{E}_n=V_0+\hbar\omega\left(n+\frac{1}{2}\right),
\end{equation}
where $n$ is again a non-negative integer. Substituting the value of $\rho_{\rm e}$ from Eq.~(\ref{RhoEq}) into $V_{\rm eff}(\rho_{\rm eq})$ and ${\rm d}^2V_{\rm eff}/{\rm d\rho^2}|_{\rho=\rho_{\rm e}}$, allows us to find the quantized energy levels:
\begin{equation}\label{ApproxHOEnergyLevelsLC}
\mathcal{E}_{n\ell}\approx\,\frac{\hbar eB}{m}\left(n+\frac{1}{2}+\frac{1}{2}\sqrt{\ell^2-\tfrac{1}{4}}-\frac{\ell}{2}\right)-\frac{amc^2}{2\sqrt{\ell^2-\tfrac{1}{4}}}\left[n+\frac{1}{2}+\sqrt{\ell^2-\tfrac{1}{4}}\ln\left(\frac{2\hbar}{eB\rho_*^2}\sqrt{\ell^2-\tfrac{1}{4}}\right)\right].
\end{equation}
This is the expression of the energy levels of the particle  --- up to the first order in $a$ --- for each quantum number $n$ and for each quantum orbital $\ell$. It should be recalled, though, that, as mentioned above, for the case of $K\neq1$ one has to replace in this result $\ell$ by $\ell/K$. We clearly see now form this expression that we recover the usual Landau levels plus the first-order correction we obtained using perturbation theory. These, of course, do agree exactly in the large-$\ell$ limit. Indeed, for large $\ell$, Eq.~(\ref{ApproxHOEnergyLevelsLC}) becomes identical to Eq.~(\ref{1ellCorrection}) obtained for the particular case $n=1$. On the other hand, for $a=0$ ({\it i.e.}, by removing the cylinder), formula (\ref{ApproxHOEnergyLevelsLC}) reproduces, in the large-$\ell$ limit, the familiar Landau energy levels of a particle inside a constant and uniform magnetic field within the Minkowski spacetime.

From the full expression (\ref{ApproxHOEnergyLevelsLC}), we also see that the first-order correction depends logarithmically on the magnetic field as well. However, in contrast to the case of the spherical mass \cite{GravityLandau}, and the Kerr spacetime case we are going to see shortly, we do not obtain quantized energy levels when putting $B=0$. In fact, when putting $B=0$ in Eq.~(\ref{ApproxHOEnergyLevelsLC}) the first line vanishes whereas the second line blows up. This is due to the fact that the equilibrium distance $\rho_e$ does not actually exist in the absence of the magnetic field as the logarithmic potential alone does not allow for any equilibrium position of the particle with $\ell\neq0$. In the absence of the magnetic field, we are left with an infinitely long cylinder the gravitational field of which is unable to counterbalance the centrifugal force on the particle due to the circular motion of the latter. The simple harmonic approximation does not therefore work for a pure gravitational field created by an infinitely long cylinder. 

It is now enlightening to examine the orders of magnitude involved in such energy levels splittings. For a magnetic field of the order of $10\,$T --- now easily achievable in a laboratory \cite{StrongB} --- and using a $1\,$cm-radius cylinder of pure platinum and $2$ meters in length for the infinitely-long cylinder approximation to hold, leads to a first-order correction to the first Landau levels of the order of $10^{-19}\,$eV. This small energy difference is, unfortunately, still too small for the presently achievable resolution which is of the order of $10^{-15}\,$eV \cite{TodayLimit}. To remedy this, one would just have to increase the size of the cylinder. In fact, using a $1$\,m-radius cylinder would effectively increase such a gravitational correction by four orders of magnitude to easily reach the present sensitivity limit of $\Delta E\sim 10^{-15}\,$eV. The only downside is that one would then have to increase the length of the cylinder accordingly.

Before we move on to the case of a particle inside the Kerr metric, we would like to note here three important facts. The first two are the ones already pointed out in Ref.~\cite{GravityLandau} for the case of the spherical mass and which still apply here. The first is that it is actually possible to rely solely on the solutions to Eq.~(\ref{KGinLC2}) and extract the energy quantization condition without making use of the Schr\"odinger equation (\ref{KGinLC3}). In fact, while it is obvious that the logarithm in Eq.~(\ref{KGinLC2}) makes the latter hardly solvable analytically, by expanding the logarithmic function one might turn the equation into a Heun-like differential equation \cite{BookHeun}. Such a differential equation has well-known solutions, called Heun functions. The procedure then consists in imposing either one of two specific conditions on such a function to guarantee the square-integrability of the latter, and hence to provide it with a physical meaning \cite{GravityLandau}. The problem with such a procedure, as explained in detail in Ref.~\cite{GravityLandau}, is that one of the conditions to impose does not provide a consistent quantization of energy for arbitrary values of the mass-source of the gravitational field, while the other condition does not allow to practically extract a simple answer as it involves finding the zeros of an infinite series. For this reason, we are not going to dwell more on these other two approaches here. 

The last point we would like to comment on here is that it would be natural now to attempt to apply the same techniques used above to the case of a rotating cylinder. Unfortunately, however, to deal with such a case one has to use the so-called Lewis spacetime \cite{Cunha}, which is even more complicated than the metric (\ref{LCMetric}). Given that the Lewis spacetime reduces in the limit of zero radius of the cylinder to that of a rotating cosmic string (see, e.g., Ref.~\cite{Bronnikov} and the references therein for more details about such a metric), which, in turn, has extensively been studied in Ref.~\cite{Lewis}, we are going to turn instead into the rotating spacetime represented by the Kerr metric. The latter is indeed much more prone to experimental verification, both at the tabletop experiments level and at the astrophysical level.

\section{A particle inside a magnetic field in the Kerr spacetime}\label{sec:III}

In this section, our test particle is still a charged spinless particle moving in the plane perpendicular to the constant and uniform magnetic field $\bf B$. Now, however, we assume the particle is going around a massive sphere of radius $r_0$, of mass $M$, and of angular momentum $J$ the direction of which is parallel to that of the magnetic field. In the weak-field approximation and slow rotation of the mass source, the Kerr metric around a rotating sphere of mass $M$ and angular momentum $J$ takes the following form in the spherical coordinates $(t,r,\theta,\phi)$ (see, e.g., Ref.~\cite{Kerr}),
\begin{equation}
\label{Kerr}
{\rm d}s^2=-\left(1-\frac{2GM}{c^2r}\right)c^2{\rm d}t^2+\left(1+\frac{2GM}{c^2r}\right){\rm d}r^2+r^2{\rm d}\Omega^2-\frac{4GM\alpha}{cr}\sin^2\theta\,{\rm d}t\,{\rm d}\phi.
\end{equation}
Here, ${\rm d}\Omega^2={\rm d}\theta^2+\sin^2\theta{\rm d}\phi^2$ and $\alpha=J/Mc$ is the specific angular momentum, that is, the angular momentum per unit mass, of the rotating sphere. To the first order in $GM/c^2r$ and in $GM\alpha/c$, at which we expanded this metric, the square root of the determinant of the metric is $\sqrt{-g}\approx cr^2\sin\theta$. On the other hand, because of the time-independence of both the gravitational field and of the magnetic field, and because of the symmetry around the $z$-axis, we expect the wavefunction of the test particle of energy $E$ to be of the form, $\varphi(t,r,\theta,\phi)=e^{-i\frac{Et}{\hbar}}e^{i\ell\phi}R(r,\theta)$. We assume again that the particle has no momentum along the $z$-direction and that it is moving counterclockwise around the sphere. Therefore, in the symmetric gauge, expressed in a covariant form in spherical coordinates, the non-vanishing components of the potential vector in a rotating spacetime read \cite{Wald}, $A_t=\frac{1}{c}B\alpha g_{tt}+\frac{1}{2}Bg_{t\phi}=-cB\alpha[1-\frac{GM}{c^2r}(2-\sin^2\theta)]$ and $A_\phi=\frac{1}{2}Bg_{\phi\phi}+\frac{1}{c}B\alpha g_{t\phi}\approx\tfrac{1}{2}Br^2\sin^2\theta.$\footnote{Note that we displayed here again the covariant form of the potential vector as the tetrad form $A_{\hat{\phi}}=\frac{1}{2}Br\sin\theta$ we used in our previous version of the manuscript leads to much confusion. The advantage of the tetrad expression is that it has the right dimensions for a potential vector and it allows one to straightforwardly extract the magnetic field from the spatial components, $\vec{\bf B}=\nabla\times\vec{\bf A}$. However, one has then to take special care when plugging such an expression inside Eq.~(\ref{KG}). The result one obtains is, of course, the same with both expressions.} Here, we have kept only the first-order in $\alpha$. The effect of the constant term in the time-component $A_t$ consists simply in redefining the energy reference of the charged particle in the spacetime by shifting the energy of the latter by the constant $-ceB\alpha$. Therefore, by redefining the energy reference for $E$ by performing the shift $E\rightarrow E+eB\alpha c$, the constant term in $A_t$ is absorbed. The Klein-Gordon equation for the particle in this curved spacetime then takes, up to the first-order in $\alpha$, the following form,
\begin{align}\label{KGinKerr1}
&\Bigg[\!\left(\frac{E^2}{\hbar^2c^2}+\frac{2GMEeB\alpha(2-\sin^2\theta)}{\hbar^2c^3r}\right)\!\left(1+\frac{2GM}{c^2r}\right)\!-\!\frac{m^2c^2}{\hbar^2}\!+\!\left(1-\frac{2GM}{c^2r}\right)\!\partial_r^2\!
+\!\left(\frac{2}{r}-\frac{2GM}{c^2r^2}\right)\!\partial_r\nonumber\\
&+\frac{\partial^2_\theta}{r^2}
+\frac{\cos\theta\partial_\theta}{r^2\sin\theta}-\frac{\ell^2}{r^2\sin^2\theta}+\frac{eB\ell}{\hbar}
-\frac{e^2B^2r^2\sin^2\theta}{4\hbar^2}-\frac{4GM\alpha E\ell}{\hbar c^3r^3\sin^2\theta}+\frac{2GM\alpha EeB}{\hbar^2c^3r}\Bigg]R(r,\theta)=0.
\end{align}  
Further, by having the particle move along the equatorial plane, along which $\theta=\tfrac{\pi}{2}$, the cylindrical symmetry of the system allows us to also expect the radial function $R(r,\theta)$ to depend only on the distance $\rho=r\sin\theta$ of the particle from the $z$-axis which is perpendicular to the plane of motion. This would then make the radial function $R(r,\theta)$ a function of the form $R(r,\sin\theta)=R(r\sin\theta)=R(\rho)$. Therefore, we can use the greatly simplifying identities, $\partial_\theta R=r\cos\theta\partial_\rho R$ and $\partial_\theta^2 R=-r\sin\theta\partial_\rho R+r^2\cos^2\theta \partial_\rho^2R$. In fact, substituting these into the previous equation, the latter takes the following simplified explicit form for $\theta=\tfrac{\pi}{2}$:
\begin{multline}\label{KGinKerr3}
\frac{{\rm d^2} R}{{\rm d}\rho^2}+\frac{1}{\rho}\frac{{\rm d}R}{{\rm d}\rho}+\Biggl[\frac{E^2}{\hbar^2c^2}\left(1+\frac{4GM}{c^2\rho}\right)\\
-\left(1+\frac{2GM}{c^2\rho}\right)\bigg(\frac{m^2c^2}{\hbar^2}+\frac{\ell^2}{\rho^2}-\frac{eB\ell}{\hbar}+\frac{e^2B^2\rho^2}{4\hbar^2}+\frac{4GME\ell\alpha}{\hbar c^3\rho^3}-\frac{4GM EeB\alpha}{\hbar^2c^3\rho}\bigg)\Biggr]R=0.
\end{multline}
By performing the change of variable $R(\rho)=\psi(\rho)/\sqrt{\rho}$, and then decomposing the energy of the test particle as, $E=\mathcal{E}+mc^2$, and using again the usual non-relativistic approximation $E^2\approx2mc^2\mathcal{E}+m^2c^4$, the above equation becomes, after keeping only the leading terms in $\alpha$ and $GM/c^2\rho$, the final Schr\"odinger equation reads,
\begin{equation}\label{SchrodinKerr}
-\frac{\hbar^2}{2m}\psi''+\Bigg[\frac{e^2B^2\rho^2}{8m}+\frac{\hbar^2(\ell^2-\tfrac{1}{4})}{2m\rho^2}-\frac{\hbar eB\ell}{2m}-\frac{GMm}{\rho}\left(1+\frac{2eB\alpha}{mc}\right)+\frac{2\hbar GM\ell\alpha}{c\rho^3}\Bigg]\psi=\mathcal{E}\psi.
\end{equation}
This equation looks very similar to Eq.~(18) of Ref.~\cite{GravityLandau} found for the weak-field limit of the Schwarzschild spacetime. The only difference is, indeed, the presence of the correction term $(1+2eB\alpha/mc)$ multiplying the Newtonian term $GMm/\rho$, as well as the extra term which decreases with the inverse cube of the distance $\rho$ of the particle from the center of the sphere. The reason we kept this latter term, despite the $\hbar$ multiplying it, is, as we shall below, the orbital number $\ell$ could become very large. In fact, it could become as large as $eB\rho_0^2/\hbar$, in which case the last term inside the square brackets of Eq.~(\ref{SchrodinKerr}) becomes of the same order as the first term to its left. Nonetheless, having already obtained the necessary tools for dealing with such extra perturbative terms inside the Schr\"odinger equation in Ref.~\cite{GravityLandau}, we can greatly benefit here from the results in that reference concerning the Newtonian term and its Kerr correction. For the last term, however, a new integral is required and is given in appendix \ref{A}. We are going therefore to apply here also the two different approaches used in Section~\ref{sec:II} to extract the quantized energy levels based on the key results given in Ref.~\cite{GravityLandau}. Note that, similarly to what we discussed in Section \ref{sec:II} for the cylinder, the wavefunctions $\psi_{n\ell}^{(0)}$ around a rotating sphere of finite radius $\rho_0$ have to satisfy the continuity condition (\ref{FContinuity}). However, we are going to simply assume here again that the radius of the sphere and the magnitude of the magnetic field are such that quantum numbers $n$ and $\ell$ are guaranteed for the particle.
\subsection{Using perturbation theory}
Treating Eq.~(\ref{SchrodinKerr}) with the time-independent perturbation theory gives in fact the perturbed Landau energy levels at the first order as \cite{GravityLandau}, $\mathcal{E}_{n\ell}=\mathcal{E}_n^{(0)}+\braket{\psi_{n\ell}^{(0)}|V(\rho)|\psi_{n\ell}^{(0)}}$, where $V(\rho)$ is the perturbing potential given by the last three terms inside the square brackets of Eq.~(\ref{SchrodinKerr}). Thus, by adopting the specific expression (18) found in Ref.~\cite{GravityLandau} --- after inserting the correcting factor $(1+2eB\alpha/mc)$ there --- and adding the contribution of the cubic term in Eq.~(\ref{SchrodinKerr}), we immediately find the first-order correction to the Landau levels as follows: 
\begin{align}\label{Reduced1stOrderSplitting}
\mathcal{E}_{n\ell}&=\mathcal{E}_{n}^{(0)}-GMm\left[\left(1+\frac{2eB\alpha}{mc}\right)\mathcal{P}_{n\ell}\mathcal{M}_{n\ell}^{-1}-\left(\frac{2\hbar\alpha}{mc}\ell\right)\mathcal{Q}_{n\ell}\mathcal{M}_{n\ell}^{-1}\right]\nonumber\\
&=\mathcal{E}_{n}^{(0)}-GMm\sqrt{\frac{eB}{2\hbar}}\left[\left(1+\frac{2eB\alpha}{mc}\right)\bar{\mathcal{P}}_{n\ell}\bar{\mathcal{M}}_{n\ell}^{-1}-\left(\frac{eB\alpha}{mc}\ell\right)\bar{\mathcal{Q}}_{n\ell}\bar{\mathcal{M}}_{n\ell}^{-1}\right].
\end{align}
The factor $\mathcal{P}_{n\ell}$ represents an infinite series and is given by expression (\ref{AppendixIntegralP}) of appendix \ref{A} after setting the integers $m=n$ there. It arises from the $1/\rho$ Newtonian term in the potential. The factor $\mathcal{Q}_{n\ell}$ is also an infinite series and is given by expression (\ref{AppendixIntegralQ}). It arises from the $1/\rho^3$ term in the potential. In the second line, we have introduced again the reduced series $\bar{\mathcal{M}}_{n\ell}$ which consists of expression (\ref{AppendixIntegralM}) but without the constant factor $(2/\beta)^{\ell+1}$. Similarly, we introduced the reduced series $\bar{\mathcal{P}}_{n\ell}$ and $\bar{\mathcal{Q}}_{n\ell}$ which consist of expressions (\ref{AppendixIntegralP}) and (\ref{AppendixIntegralQ}), respectively, in which we suppress the constant factors $(2/\beta)^{\ell+\frac{1}{2}}$ and $(2/\beta)^{\ell-\frac{1}{2}}$, respectively. This, in fact, allows us to get explicitly the factor $\sqrt{eB/2\hbar}$ out in Eq.~(\ref{Reduced1stOrderSplitting}). This result shows how the degenerate Landau levels split at the first-order in $GMm$ due to gravity. The ratio $eB\alpha/mc$ contains the frame-dragging effect created by the rotating curved spacetime. 

In order to appreciate this result, it is instructive to examine the fate of the first Landau level $n=1$. On the one hand, according to the definitions (\ref{AppendixIntegralMDef}), (\ref{AppendixIntegralPDef}) and (\ref{AppendixIntegralQDef}), for small values of $\ell$, the product $\mathcal{P}_{1\ell}\mathcal{M}_{1\ell}^{-1}$ is simply of the order of $1/\rho_0$ whereas the product $\mathcal{Q}_{1\ell}\mathcal{M}_{1\ell}^{-1}$ is of the order of $1/\rho^3_0$.  The last term in the first line in Eq.~(\ref{Reduced1stOrderSplitting}) then becomes suppressed simply because of the presence of the $\hbar$ factor in the numerator. The splitting of the Landau levels in this case reduces to the following first-order correction,
\begin{equation}\label{n=1SplittingEllSmall}
\mathcal{E}_{1\ell}\approx\frac{3\hbar eB}{2m}-\frac{GMm}{\rho_0}\left(1+\frac{2eB\alpha}{mc}\right).
\end{equation}
On the other hand, for large values of $\ell$, substituting the large-$\ell$ limits (\ref{M1ellInfinite}), (\ref{P1ellInfinite}) and (\ref{Q1ellInfinite}) of $\mathcal{M}_{1\ell}$, $\mathcal{P}_{1\ell}$ and $\mathcal{Q}_{1\ell}$, respectively, inside Eq.~(\ref{Reduced1stOrderSplitting}) we find, after using the asymptotic property of the gamma function $\Gamma(z)\sim z^{-\frac{1}{2}}e^{z(\log z-1)}$ \cite{BookKummer}, the following approximation for the energy splitting of the first Landau level:
\begin{equation}\label{n=1SplittingEllInfinite}
\mathcal{E}_{1,\ell\gg1}\approx\frac{3\hbar eB}{2m}-GMm\sqrt{\frac{eB}{2\hbar\ell}}\left(1+\frac{eB\alpha}{mc}\right).
\end{equation}
We see that the Newtonian correction term in Eq.~(\ref{n=1SplittingEllInfinite}) decreases like $1/\sqrt{\ell}$, and therefore becomes gradually suppressed for large $\ell$ as it was the case for a static spherical mass \cite{GravityLandau}. In addition, the frame-dragging correcting factor itself does not depend on the orbital quantum number $\ell$. In contrast, for large $n$ we see from Eqs.~(\ref{AppendixIntegralM}), (\ref{AppendixIntegralP}) and (\ref{AppendixIntegralQ}), giving $\mathcal{M}_{n\ell}$, $\mathcal{P}_{n\ell}$ and $\mathcal{Q}_{n\ell}$, respectively, that the first-order correction (\ref{Reduced1stOrderSplitting}) does not decrease with an increasing $n$. On the other hand, for large $\ell$, we see from Eq.~(\ref{n=1SplittingEllInfinite}) that the correction becomes for $n=1$ insensitive to the radius $\rho_0$ of the rotating sphere. The same remark is valid for $n>1$, though.

It should be kept in mind here, as emphasized in Ref.~\cite{GravityLandau}, that the large-$\ell$ approximation obtained in Eq.~(\ref{n=1SplittingEllInfinite}) is valid for very large values of $\ell$. This is because the large-$\ell$ limit in the appendix was found by taking into account the very large term $\frac{\beta}{2}\rho_0^2$ appearing inside the incomplete gamma functions in Eqs.~(\ref{M1ell}), (\ref{P1ell}) and (\ref{Q1ell}). As a consequence, and contrary to what it might seem at first sight, the correction term obtained on the right-hand side in Eq.~(\ref{n=1SplittingEllInfinite}) is really small as is required for a perturbation.

Similarly, using the results of Ref.~\cite{GravityLandau} for the second order in the perturbation theory, we can also easily deduce the second-order correction to the energy levels. However, given the already complicated first-order expression (\ref{Reduced1stOrderSplitting}), we are not going to display the second order-correction here, suffice it to note that it is going to be quadratic in $GMm$ as was the case for the static spherical mass. The frame-dragging effect will then simply appear as corrections terms proportional to various powers of the ratio $eB\alpha/mc$.
\subsection{Using the harmonic oscillator approximation}
Let us now apply here the method based on approximating the effective potential of the particle by that of a simple harmonic oscillator. Unfortunately, the presence of the last term inside the effective potential in the Schr\"odinger equation (\ref{SchrodinKerr}) renders this method analytically intractable for arbitrary values of the orbital number $\ell$. In fact, the condition ${\rm d}V_{\rm eff}/{\rm d}\rho=0$, that would give the radius $\rho_e$ at which the potential reaches its minimum, becomes in this case a quintic equation. For this reason, the harmonic oscillator approximation becomes really useful only for small values of $\ell$, for then the last term in the effective potential in Eq.~(\ref{SchrodinKerr}) can be neglected. 

Therefore, given that for the case of small $\ell$ the only difference between the Schr\"odinger equation of our system as given by Eq.~(\ref{SchrodinKerr}) and that of Ref.~\cite{GravityLandau} resides only in the correcting factor $(1+2eB\alpha/mc)$ that multiplies the Newtonian potential, we are not going to display here the details of the calculations. We are going to content ourselves by displaying the final results after inserting such a correcting term. In addition, since within the perturbation theory we used above we restricted ourselves to the first-order approximation, we are not going to display the second-order correction here either.

Based on the general formula for the perturbed energy levels in the spherical static mass \cite{GravityLandau}, the energy levels for the rotating mass thus split at the first order in the specific angular momentum $\alpha$ as follows:
\begin{align}\label{ApproxHOEnergyLevels}
\mathcal{E}_{n\ell}&\approx\,\frac{\hbar eB}{m}\left(n+\frac{1}{2}+\frac{1}{2}\sqrt{\ell^2-\frac{1}{4}}-\frac{\ell}{2}\right)\nonumber\\&\quad+\frac{GMm}{(\ell^2-\tfrac{1}{4})^{3/4}}\sqrt{\frac{eB}{32\hbar}}\left(1+\frac{2eB\alpha}{mc}\right)\left(n+\frac{1}{2}-4\sqrt{\ell^2-\frac{1}{4}}\right).
\end{align}
We clearly see form this result that we recover again the usual Landau levels plus a similar formal structure for the first-order correction we obtained using perturbation theory. The dependence of the correction on the square root of the magnetic field and on the ratio $eB\alpha/mc$ is remarkable. Of course, despite these similarities between the results of the two methods at this first-order level, the result (\ref{ApproxHOEnergyLevels}) cannot be used for large values of $\ell$, in contrast to the result (\ref{n=1SplittingEllInfinite}) which is specifically found for large $\ell$. This particular case shows the superiority in this investigation of the approach based on perturbation theory over the simple harmonic oscillator approximation.

As was the case with the results obtained in Ref.~\cite{GravityLandau} concerning the static spherical mass, our results here for the rotating spherical mass might {\it a priori} both be applied at the tabletop experiments level and at the astrophysical observations level. Unfortunately, as we shall see, for the latter case our above approximations become too restrictive to be applicable for the wide range of astrophysical situations. In fact, our approximation does show that for the Landau quantization to be significant, the frame-dragging contribution to the effective potential of the particles should not dominate the interaction of the latter with the magnetic field. 

Indeed, with protons as the test particles, a $1$\,m-radius spherical mass of platinum, and a laboratory magnetic field of the order of $10\,$T, the first-order correction to the first Landau levels of the protons is, according either to Eq.~(\ref{n=1SplittingEllSmall}) or Eq.~(\ref{n=1SplittingEllInfinite}), of the order of $10^{-8}\,$eV. If the sphere is then rotated at about $100$ revolutions per minute, the frame-dragging effect induces the dimensionless correction to the Newtonian potential, $eB\alpha/mc$, which is of the order of $10^{-7}$. For electrons, this dimensionless factor would be of the order of $10^{-4}$. Of course, due to the presence of the magnetic field, the rotating platinum spherical mass should be grounded in order to avoid any induced electric current. 

On the other hand, at the astrophysical level, it is already known in the literature that the strong magnetic fields around rotating neutron stars, magnetars and magnetic white dwarfs could be taken into account to study how the equations of states of the surface (or even the bulk) nuclei matter would be affected by the Landau quantization caused by such strong magnetic fields \cite{Broderick,Chamel,StarsBook}. However, these astrophysical objects could acquire, in addition to the strong magnetic fields, very high rotational speeds that could reach up to $10^4$ revolutions per minute. The contribution to the splitting of the energy levels in Eqs.~(\ref{n=1SplittingEllSmall}) and (\ref{n=1SplittingEllInfinite}) due to the frame-dragging effect becomes then dominant over the contribution due to the Newtonian potential and even over the Landau energy levels themselves. For a $10\,$kilometer-radius neutron star, rotating at such a rate and producing a magnetic field of the order of $10^{10}\,$T, which is also typical of magnetars \cite{McGillCatalogue}, the frame-dragging term $eB\alpha/mc$ is already of the order $10^{15}$ for electrons and of the order of $10^{11}$ for protons. Our weak-field approximation due to a slow rotation of the mass source then breaks down in this case. Actually, such strong magnetic fields combined with a radius of the star that is above one kilometer keeps the frame-dragging effect dominant unless the rotation rate of the star is much smaller than one revolution per year.

\section{Discussion \& Conclusion}\label{sec:V}

We have studied the effect of two different gravitational fields on a charged particle moving inside a uniform and constant magnetic field. The first consists of the field created by an infinitely long cylinder, expressed in the form of the Levi-Civita metric, and the second one was the field created by a rotating spherical mass, expressed in the form of the Kerr spacetime. We found that the infinite Landau degeneracy is removed in both cases as the Landau orbitals of the same Landau level split in energy. As was done in Ref.~\cite{GravityLandau} for the Schwarzschild spacetime case, we used here two independent methods to reach the quantized energy levels implied by the corresponding curved-spacetime Klein-Gordon equations. 

The results of the two methods are quantitatively different due to the different degrees of approximation each method relies on. Both methods, however, point towards the same qualitative splitting of the energy levels. In the case of the Levi-Civita metric the splitting is characterized by a logarithmic dependence on the radius of the cylinder and of the radius of the position taken as a reference for the gravitational potential. Our results for this metric would be valid in a realistic setup provided one uses a very long and very thin massive cylinder, with the test particle moving very closely to the surface of the cylinder. This first investigation is more of a gravitational-testing tool. It provides an additional important approach towards testing the century-old and apparently illusive Levi-Civita metric.

The second investigation provided us with a very nice way of testing the famous frame-dragging of general relativity at the level of quantum particles. The larger the specific angular momentum of the rotating massive sphere is, the bigger is the splitting in the energy of the Landau levels. This second investigation is testable at the level of tabletop experiments using strong magnetic fields and rapidly rotating massive grounded spheres. Both investigations have been carried out using, for simplicity, spinless particles. Such a setup can indeed easily be achieved experimentally by using heavy ions the total spin of which is negligible. 

At the level of astrophysical observations of rapidly rotating neutron stars, magnetars and magnetic white dwarfs, our investigation showed that for a wide range of realistic astrophysical objects (with fast rotations and strong magnetic fields) the frame-dragging effect cannot constitute a mere perturbation compared to the Newtonian potential neither compared to the Landau levels themselves. We saw that the frame-dragging effect couples to the magnetic field in such a way that the effect of the latter alone on the particles is what actually constitutes a perturbation. Therefore, because of the frame-dragging effect the Landau levels would emerge and dominate on such highly magnetized stars only when the latter are slowly rotating around their axes.

We have based our whole approach in this paper on the combination of the Klein-Gordon equation in curved spacetime and the full spacetime metrics of both the Kerr and Levi-Civita spacetimes. The full equations (\ref{KGinLC2}) and (\ref{KGinKerr3}) have then been approximated into much easier to solve equations by relying on the low-curvature and non-relativistic regime approximations. Such restrictions have been dictated by, respectively, the orders of magnitude of the massive sources and of the magnetic fields in which we are interested in this paper. Our main goal in this paper has indeed been to simply bring into light the effect of more complicated gravitational fields than that due to a static spherical mass on the Landau quantum levels. A fully relativistic treatment of the motion of charged particles in a strong magnetic field and in a curved spacetime, as done in, e.g., Refs.~{\cite{Magnetized1,Magnetized2,Magnetized3,Magnetized4,Magnetized5,Magnetized6,Magnetized7,Magnetized8,Magnetized9,Magnetized10}}, will be the next step. We defer the investigation taking into account the relativistic corrections to the motion of the electrons or neutrons moving on the surface of neutron stars/magnetars/magnetic white dwarfs to forthcoming works. We shall then conduct rigorously a detailed study of the fate of the equation of state on these astrophysical objects caused by the splitting of the Landau levels due to their rotation. In fact, on the one hand, going beyond the non-relativistic regime leads to extra terms of the form $\rho^2\ln\rho$ inside Eq.~(\ref{KGinLC2}) for the Levi-Civita spacetime and might allow one to get to higher order approximations in the parameters $a$ and $b$ of the Levi-Civita metric. On the other hand, allowing for a relativistic regime of the test particle would lead to non-perturbative terms of the form $1/\rho$ and $1/\rho^3$ inside Eq.~(\ref{KGinKerr3}) for the Kerr spacetime. The presence of all these extra terms necessitates different mathematical methods for solving the corresponding differential equations than those adopted here.

\section*{Acknowledgments}
We are grateful to 
Bobur Turimov for the helpful comment about our notation for the potential vector and for having pointed out to us Ref.~\cite{Wald} that contains the complete form of the potential vector in rotating spacetimes. We are also grateful to the anonymous referee for the pertinent comment that led us to rectify a previous erroneous version of the condition (\ref{FContinuity}) imposed on the wavefunction. This work is supported by the Natural Sciences and Engineering Research
Council of Canada (NSERC) Discovery Grant (RGPIN-2017-05388).


\appendix
\section{Evaluating integrals involving products of Kummer's functions, power functions and a logarithm}\label{A}
In this appendix we give the results for the integrals needed in the text and give brief outlines of their derivation, referring for more details to Ref.~\cite{GravityLandau}.
The various integrals needed are improper integrals involving the product of two Kummer's functions, powers of the distance $\rho$ from the center of motion and a logarithm involving the distance $\rho$. For this purpose we need to recall the following general result from Ref.~\cite{GravityLandau}, which involves an integral of two Kummer's functions \cite{BookKummer} with an arbitrary power-function $x^{c-1}$:
\begin{multline}\label{AppendixKummerIntegralOffset}
\int_{x_0}^{\infty}x^{c-1}e^{-zx}\,_1F_1(-n,b+1;zx)\,_1F_1(-m,b+1;zx){\rm d}x\\
=\sum\limits_{k=0}^{\infty}\sum\limits_{\substack{p=0\\q=0}}^\infty\frac{e^{-zx_0}\Gamma(c)(-n)_p(-m)_q\,x_0^{p+q+k}}{\,k!\,p!\,q!\,(b+1)_p(b+1)_q}\frac{\Gamma(b+1+q)\Gamma(m+b+1+k-c)}{\Gamma(b+q+1+k-c)\Gamma(m+b+1)}\,z^{p+q+k-c}\\
\times\,_3F_2(-n+p,c-k,c-k-b-q; b+p+1,c-k-m-b;1).
\end{multline}
Here, the symbol $(a)_k$ stands for the product $(a)_k=a(a+1)\ldots(a+k-1)$, such that, by definition, $(a)_0=1$. It is often called in the literature the Pochhammer symbol \cite{BookKummer}. The special functions $\,_3F_2(a,b,c;d,e;1)$ are the so-called generalized hypergeometric functions \cite{BookHyperGeometric} and $\Gamma(x)$ is the gamma function \cite{BookKummer}.
First, the expression of $\mathcal{M}_{n\ell}$, needed to find the normalization constants $A_{n\ell}$ in Section~\ref{sec:II}, is based on the normalization condition,
\begin{align}
\int_{\rho_0}^{\infty}\psi^{(0)*}_{n\ell}(\rho)\psi^{(0)}_{n\ell}(\rho)\,{\rm d}\rho=1,\nonumber
\end{align}
involving the unperturbed wavefunctions $\psi_{n\ell}^{(0)}(\rho)$ given explicitly by Eq.~(\ref{UnperturbedWF}). Using the general result (\ref{AppendixKummerIntegralOffset}), we can evaluate the left-hand side of this integral by performing the change of variable, $x=\rho^2$, and setting $b=\ell$, $c=\ell+1$, $z=\beta/2$ and $x_0=\rho_0^2$. The explicit expression of $\mathcal{M}_{mn\ell}$, from which the needed quantities $\mathcal{M}_{n\ell}$ can be extracted by setting $n=m$, is found to be \cite{GravityLandau},
\begin{equation}\label{AppendixIntegralMDef}
\int_{\rho_0}^{\infty}\rho^{2\ell+1}e^{-\frac{\beta}{2}\rho^2}\,_1F_1\left(-n,\ell+1;\frac{\beta}{2}\rho^2\right)\,_1F_1\left(-m,\ell+1;\frac{\beta}{2}\rho^2\right)\,{\rm d}\rho=\mathcal{M}_{mn\ell},
\end{equation}
where,
\begin{multline}\label{AppendixIntegralM}
\mathcal{M}_{mn\ell}=\sum\limits_{k=0}^{\ell}\sum\limits_{\substack{p=0\\q=0}}^\infty\frac{e^{-\frac{\beta}{2}\rho_0^2}\Gamma(\ell+1)(-n)_p(-m)_q\,\rho_0^{2(p+q+k)}}{2\,k!\,p!\,q!\,(\ell+1)_p(\ell+1)_q}\frac{\Gamma(\ell+1+q)\Gamma(m+k)}{\Gamma(q+k)\Gamma(m+\ell+1)}\left(\frac{\beta}{2}\right)^{p+q+k-\ell-1}\\
\times\,_3F_2(-n+p,\ell+1-k,1-k-q; \ell+p+1,1-k-m;1).
\end{multline}
Notice that, in contrast to the general series (\ref{AppendixKummerIntegralOffset}), the series in $k$ in expression (\ref{AppendixIntegralM}) terminates at $k=\ell$, for in this case the exponent $c$ in Eq.~(\ref{AppendixKummerIntegralOffset}), coming from a binomial expansion (see the appendix of Ref.~\cite{GravityLandau}), is an integer.
\subsection{Integral needed in Section~\ref{sec:II}}
The integral in Section~\ref{sec:II}, involving the product of a logarithm and the unperturbed wavefunctions (\ref{UnperturbedWF}), has the form,
\begin{align}
\int_{\rho_0}^{\infty} \ln\left(\frac{\rho}{\rho_*}\right)\psi^{(0)*}_{n\ell}(\rho)\psi^{(0)}_{n\ell}(\rho)\,{\rm d}\rho,\nonumber
\end{align}
 and can be computed using the general result (\ref{AppendixKummerIntegralOffset}) by making the change of variable $x=\rho^2$, and then setting $z=\beta/2$, $b=\ell$, $c=\ell+s+1$ and $x_0=\rho_0^2$. In fact, using the identity $\ln\rho=\left(\frac{{\rm d}}{{\rm d}s}\rho^s\right)_{s=0}$ allows us to transform the above integral with a logarithm into an integral with a power function of $\rho$ to which the general result (\ref{AppendixKummerIntegralOffset}) can be applied. Doing so, we find the following result:
\begin{equation}\label{AppendixIntegralLDef}
\int_{\rho_0}^{\infty}\rho^{2\ell+1}e^{-\frac{\beta}{2}\rho^2}\ln\left(\frac{\rho}{\rho_*}\right)\,_1F_1\left(-n,\ell+1;\frac{\beta}{2}\rho^2\right)\,_1F_1\left(-m,\ell+1;\frac{\beta}{2}\rho^2\right){\rm d}\rho=\mathcal{L}_{mn\ell},\\
\end{equation}
where,\\
\begin{align}\label{AppendixIntegralL}
\mathcal{L}_{mn\ell}&=\frac{1}{2}\Bigg[\frac{\rm d}{{\rm d} s}\int_{\rho_0}^{\infty}\rho^{2\ell+2s+1}e^{-\frac{\beta}{2}\rho^2}\,_1F_1\left(-n,\ell+1;\frac{\beta}{2}\rho^2\right)\,_1F_1\left(-m,\ell+1;\frac{\beta}{2}\rho^2\right){\rm d}\rho\Bigg]_{s=0}\nonumber\\
&\quad-\int_{\rho_0}^{\infty}\rho^{2\ell+1}e^{-\frac{\beta}{2}\rho^2}\ln\rho_*\,_1F_1\left(-n,\ell+1;\frac{\beta}{2}\rho^2\right)\,_1F_1\left(-m,\ell+1;\frac{\beta}{2}\rho^2\right){\rm d}\rho\nonumber\\
&=\frac{1}{2}\left(\frac{\rm d}{{\rm d}s}\mathcal{D}_{mn,\ell+s}\right)_{s=0}-\mathcal{M}_{mn\ell}\ln\rho_*,
\end{align}
with,
\begin{multline}\label{AppendixIntegralD}
\mathcal{D}_{mn,\ell+s}=\sum\limits_{k=0}^{\ell+s}\sum\limits_{\substack{p=0\\q=0}}^\infty\frac{e^{-\frac{\beta}{2}\rho_0^2}\Gamma(\ell+s+1)(-n)_p(-m)_q\,\rho_0^{2(p+q+k)}}{2\,k!\,p!\,q!\,(\ell+1)_p(\ell+1)_q}\\
\times\frac{\Gamma(\ell+q+1)\Gamma(m+k-s)}{\Gamma(q+k-s)\Gamma(m+\ell+1)}\,\left(\frac{\beta}{2}\right)^{p+q+k-\ell-s-1}\\
\times\,_3F_2(-n+p,\ell+s-k+1,s-k-q+1;\ell+p+1,s-k-m+1;1).
\end{multline}
\subsection{Integrals needed in Section~\ref{sec:III}}
One of the two integrals needed in Section \ref{sec:III} and involving the unperturbed wavefunctions (\ref{UnperturbedWF}), has the form,
\begin{align}
\int_{\rho_0}^{\infty} \rho^{-1}\psi^{(0)*}_{n\ell}(\rho)\psi^{(0)}_{n\ell}(\rho)\,{\rm d}\rho.\nonumber
\end{align}
To evaluate this integral we use again the general result Eq.~(\ref{AppendixKummerIntegralOffset}) after performing the change of variable $x=\rho^2$, and by setting $b=\ell$, $z=\beta/2$, $c=\ell+\frac{1}{2}$ and $x_0=\rho_0^2$. The result is the following \cite{GravityLandau}:
\begin{equation}\label{AppendixIntegralPDef}
\int_{\rho_0}^{\infty}\rho^{2\ell}e^{-\frac{\beta}{2}\rho^2}\,_1F_1\left(-n,\ell+1;\frac{\beta}{2}\rho^2\right)\,_1F_1\left(-m,\ell+1;\frac{\beta}{2}\rho^2\right)\,{\rm d}\rho=\mathcal{P}_{mn\ell},
\end{equation}
where,
\\
\begin{multline}\label{AppendixIntegralP}
\mathcal{P}_{mn\ell}=\sum\limits_{k=0}^{\infty}\sum\limits_{\substack{p=0\\q=0}}^\infty\frac{e^{-\frac{\beta}{2}\rho_0^2}\Gamma(\ell+\frac{1}{2})(-n)_p(-m)_q\,\rho_0^{2(p+q+k)}}{2\,k!\,p!\,q!\,(\ell+1)_p(\ell+1)_q}\\
\times\frac{\Gamma(\ell+q+1)\Gamma(m+\frac{1}{2}+k)}{\Gamma(q+k+\frac{1}{2})\Gamma(m+\ell+1)}\left(\frac{\beta}{2}\right)^{p+q+k-\ell-\frac{1}{2}}\\
\times\,_3F_2(-n+p,\ell+\tfrac{1}{2}-k,\tfrac{1}{2}-k-q; \ell+p+1,\tfrac{1}{2}-k-m;1).
\end{multline}
The second integral needed in Section \ref{sec:III}, has the following form:
\begin{align}
\int_{\rho_0}^{\infty} \rho^{-3}\psi^{(0)*}_{n\ell}(\rho)\psi^{(0)}_{n\ell}(\rho)\,{\rm d}\rho,\nonumber
\end{align}
The evaluation of such an integral proceeds in a similar fashion as with the previous integrals. After using the general result (\ref{AppendixKummerIntegralOffset}), performing the change of variable, $x=\rho^2$, and setting $b=\ell$, $z=\beta/2$, $c=\ell-\frac{1}{2}$ and $x_0=\rho_0^2$, we find, 
\begin{equation}\label{AppendixIntegralQDef}
\int_{\rho_0}^{\infty}\rho^{2\ell-2}e^{-\frac{\beta}{2}\rho^2}\,_1F_1\left(-n,\ell+1;\frac{\beta}{2}\rho^2\right)\,_1F_1\left(-m,\ell+1;\frac{\beta}{2}\rho^2\right)\,{\rm d}\rho=\mathcal{Q}_{mn\ell},
\end{equation}
where,
\begin{multline}\label{AppendixIntegralQ}
\mathcal{Q}_{mn\ell}=\sum\limits_{k=0}^{\infty}\sum\limits_{\substack{p=0\\q=0}}^\infty\frac{e^{-\frac{\beta}{2}\rho_0^2}\Gamma(\ell-\frac{1}{2})(-n)_p(-m)_q\,\rho_0^{2(p+q+k)}}{2\,k!\,p!\,q!\,(\ell+1)_p(\ell+1)_q}\\
\times\frac{\Gamma(\ell+q+1)\Gamma(m+k+\frac{3}{2})}{\Gamma(q+k+\frac{3}{2})\Gamma(m+\ell+1)}\left(\frac{\beta}{2}\right)^{p+q+k-\ell+\frac{1}{2}}\\
\times\,_3F_2(-n+p,\ell-\tfrac{1}{2}-k,-\tfrac{1}{2}-k-q; \ell+p+1,-\tfrac{1}{2}-k-m;1).
\end{multline}
\subsection{Computation of $\mathcal{M}_{1\ell}$, $\mathcal{L}_{1\ell}$, $\mathcal{P}_{1\ell}$ and $\mathcal{Q}_{1\ell}$}
Now, although these various resulting expressions are lengthy and cumbersome, they actually become much simpler in special cases. For concreteness, we are going to find the expressions of the quantities $\mathcal{M}_{1\ell}$, $\mathcal{L}_{1\ell}$, $\mathcal{P}_{1\ell}$ and $\mathcal{Q}_{1\ell}$ as these are needed to find the splitting of the first Landau level $n=1$. However, instead of substituting directly $m=n=1$ in the final formulas (\ref{AppendixIntegralM}), (\ref{AppendixIntegralL}), (\ref{AppendixIntegralD}) and (\ref{AppendixIntegralQ}), it is much easier actually to evaluate these directly from their integral definitions (\ref{AppendixIntegralMDef}), (\ref{AppendixIntegralLDef}), (\ref{AppendixIntegralPDef}) and (\ref{AppendixIntegralQDef}), respectively. The expressions of $\mathcal{M}_{1\ell}$ and $\mathcal{P}_{1\ell}$ have already been derived in Ref.~\cite{GravityLandau}, so we just reproduce them here for reference. The expression of $\mathcal{Q}_{1\ell}$ has not been given in Ref.~\cite{GravityLandau}, but its derivation proceeds in a similar fashion to the derivation of $\mathcal{M}_{1\ell}$ and $\mathcal{P}_{1\ell}$ given in Ref.~\cite{GravityLandau}. We find,
\begin{equation}\label{M1ell}
\mathcal{M}_{1\ell}=\frac{2^{\ell+1}}{2\beta^{\ell+1}}\Bigg[\Gamma(\ell+1,\tfrac{\beta}{2}\rho_0^2)-\frac{\Gamma(\ell+2,\tfrac{\beta}{2}\rho_0^2)}{(\ell+1)/2}+\frac{\Gamma(\ell+3,\tfrac{\beta}{2}\rho_0^2)}{(\ell+1)^2}\Bigg],
\end{equation}
\begin{equation}\label{P1ell}
\mathcal{P}_{1\ell}=\frac{2^{\ell-\tfrac{1}{2}}}{\beta^{\ell+\frac{1}{2}}}\Bigg[\Gamma(\ell+\tfrac{1}{2},\tfrac{\beta}{2}\rho_0^2)-\frac{\Gamma(\ell+\tfrac{3}{2},\tfrac{\beta}{2}\rho_0^2)}{(\ell+1)/2}+\frac{\Gamma(\ell+\tfrac{5}{2},\tfrac{\beta}{2}\rho_0^2)}{(\ell+1)^2}\Bigg].
\end{equation}
\begin{equation}\label{Q1ell}
\mathcal{Q}_{1\ell}=\frac{2^{\ell-\tfrac{1}{2}}}{2\beta^{\ell-\frac{1}{2}}}\Bigg[\Gamma(\ell-\tfrac{1}{2},\tfrac{\beta}{2}\rho_0^2)-\frac{\Gamma(\ell+\tfrac{1}{2},\tfrac{\beta}{2}\rho_0^2)}{(\ell+1)/2}+\frac{\Gamma(\ell+\tfrac{3}{2},\tfrac{\beta}{2}\rho_0^2)}{(\ell+1)^2}\Bigg].
\end{equation}
Here, the function $\Gamma(a,x)$ is the so-called incomplete gamma function \cite{BookKummer}. In the specific form used here, it is coming from the following integral \cite{GravityLandau}:
\begin{equation}\label{IncompleteGamma}
\int_{\rho_0}^{\infty}\rho^{2\sigma-1}e^{-\frac{\beta}{2}\rho^2}{\rm d}\rho=\frac{2^{\sigma-1}}{\beta^\sigma}\Gamma(\sigma,\tfrac{\beta}{2}\rho_0^2)=\frac{2^{\sigma-1}}{\beta^\sigma}\left[\Gamma(\sigma)-\sum\limits_{k=0}^{\infty}\frac{(-1)^k}{k!}\frac{\left(\frac{\beta}{2}\rho_0^2\right)^{\sigma+k}}{\sigma+k}\right]
\end{equation}
In the second step we have used the infinite series definition of the incomplete gamma function \cite{BookKummer} in order to be able, shortly, to compute its first derivative with respect to the auxiliary argument $s$. 

Now, for the expression of $\mathcal{L}_{1\ell}$, we easily find what it is explicitly given by following the same steps performed in the appendix of Ref.~\cite{GravityLandau} to get $\mathcal{M}_{1\ell}$ and $\mathcal{P}_{1\ell}$. Therefore, we only outline here the derivation and we refer to the appendix A of Ref.~\cite{GravityLandau} for more details. 

Starting from the first integral in Eq.~(\ref{AppendixIntegralL}), we use the infinite series definition of Kummer's function, $\,_1F_1(a,b;z)=\sum_{k=0}^{\infty}\frac{(a)_k}{(b)_k}\frac{z^k}{k!}$ \cite{BookKummer} in order to display the few non-vanishing terms in the unique function $\,_1F_1(-1,\ell+1;\frac{\beta}{2}\rho_0^2)$ inside the integral. Then, the main steps of the derivation are as follows:
\begin{align}\label{L1ellSigma}
\mathcal{D}_{1,\ell+s}&=\int_{\rho_0}^{\infty}\rho^{2\ell+2s+1}e^{-\frac{\beta}{2}\rho^2}\,\left[_1F_1\left(-n,\ell+1;\frac{\beta}{2}\rho^2\right)\right]^2{\rm d}\rho\nonumber\\
&=\int_{\rho_0}^{\infty}\rho^{2\ell+2s+1}e^{-\frac{\beta}{2}\rho^2}\left[1-\frac{\beta}{2(\ell+1)}\rho^2\right]^2{\rm d}\rho\nonumber\\
&=\beta^{-1}\left(\frac{2}{\beta}\right)^{\ell+s}\Bigg[\Gamma(\ell+s+1,\tfrac{\beta}{2}\rho_0^2)-\frac{\Gamma(\ell+s+2,\tfrac{\beta}{2}\rho_0^2)}{(\ell+1)/2}+\frac{\Gamma(\ell+s+3,\tfrac{\beta}{2}\rho_0^2)}{(\ell+1)^2}\Bigg].
\end{align}
From this result, we may now find the expression of the required derivative in Eq.~(\ref{AppendixIntegralL}). For that purpose, we use the infinite series definition of the incomplete gamma function, as given by the square brackets in the second line of Eq.~(\ref{IncompleteGamma}). We also use the identity $\Gamma'(x)=\Gamma(x)\Psi(x)$ \cite{BookKummer},
linking the derivative of the gamma function with respect to its argument to the so-called di-gamma function $\Psi(x)$. The latter function satisfies the recurrence relation $\Psi(x+1)=\Psi(x)+1/x$. This recurrence relation will be useful to us shortly for finding the large-$\ell$ limit of our final expression. The latter is found, after a lengthy but straightforward calculation, to be, 
\begin{multline}\label{L1ell}
\left(\frac{\rm d}{{\rm d}s}\mathcal{D}_{1,\ell+s}\right)_{s=0}=\frac{1}{2}\left(\frac{2}{\beta}\right)^{\ell+1}\Bigg\{\ln\left(\frac{2}{\beta}\right)\left[\Gamma(\ell+1,\tfrac{\beta}{2}\rho_0^2)-\frac{\Gamma(\ell+2,\tfrac{\beta}{2}\rho_0^2)}{(\ell+1)/2}+\frac{\Gamma(\ell+3,\tfrac{\beta}{2}\rho_0^2)}{(\ell+1)^2}\right]\\
+\Gamma(\ell+1)\Psi(\ell+1)-\frac{\Gamma(\ell+2)\Psi(\ell+2)}{(\ell+1)/2}+\frac{\Gamma(\ell+3)\Psi(\ell+3)}{(\ell+1)^2}\\
-\sum_{k=0}^{\infty}\frac{(-1)^k}{k!}\left(\frac{\beta\rho_0^2}{2}\right)^{\ell+k+1}\Bigg[\frac{\ln\left(\frac{\beta}{2}\rho_0^2\right)}{\ell+k+1}-\frac{\beta\rho_0^2\ln\left(\frac{\beta}{2}\rho_0^2\right)}{(\ell+k+2)(\ell+1)}\\+\frac{\left(\frac{\beta}{2}\rho_0^2\right)^2\ln\left(\frac{\beta}{2}\rho_0^2\right)}{(\ell+k+3)(\ell+1)^2}-\frac{1}{(\ell+k+1)^2}
+\frac{\beta\rho_0^2}{(\ell+1)(\ell+k+2)^2}-\frac{\left(\frac{\beta}{2}\rho_0^2\right)^2}{(\ell+1)^2(\ell+k+3)^2}\Bigg]\Bigg\}.
\end{multline}
As we can see, all these three expressions of $\mathcal{M}_{1\ell}$, $\mathcal{P}_{1\ell}$ and $\left(\frac{\rm d}{{\rm d}s}\mathcal{D}_{1,\ell+s}\right)_{s=0}$ are long and cumbersome. It is, however, possible, and very instructive, to find an estimate for these quantities for large $\ell$-limits for which all three of them simplify indeed greatly and allow us to have a physical idea of the energy levels splitting in Sections~\ref{sec:II} and \ref{sec:III}. 

As for $\mathcal{M}_{1\ell}$ and $\mathcal{P}_{1\ell}$, we already found their explicit asymptotic expressions for $\ell\gg1$ in Ref.~\cite{GravityLandau}. The asymptotic expression of $\mathcal{Q}_{1\ell}$ is found here in a similar way. After using the property $\Gamma(x+1)=x\Gamma(x)$ of the gamma function \cite{BookKummer}, the asymptotic expressions read, respectively,
\begin{equation}\label{M1ellInfinite}
\mathcal{M}_{1,\ell\gg1}\approx\frac{2^{\ell}}{\beta^{\ell+1}}\frac{\Gamma(\ell+1)}{\ell+1},
\end{equation}
\begin{equation}\label{P1ellInfinite}
\mathcal{P}_{1,\ell\gg1}\approx\frac{2^{\ell-\frac{1}{2}}}{\beta^{\ell+\frac{1}{2}}}\frac{(\ell+\tfrac{3}{4})\Gamma(\ell+\tfrac{1}{2})}{(\ell+1)^2},
\end{equation}
\begin{equation}\label{Q1ellInfinite}
\mathcal{Q}_{1,\ell\gg1}\approx\frac{2^{\ell-\frac{3}{2}}}{\beta^{\ell-\frac{1}{2}}}\frac{(\ell+\tfrac{7}{4})\Gamma(\ell-\tfrac{1}{2})}{(\ell+1)^2}.
\end{equation}
In a similar fashion, based on identical steps, we find here the following additional asymptotic expression,
\begin{equation}\label{DerivativeL1ellInfinite}
\frac{1}{2}\left(\frac{\rm d}{{\rm d}s}\mathcal{D}_{1,\ell+s}\right)_{s=0,\ell\gg1}
\approx\frac{1}{2}\left[\ln\left(\frac{2}{\beta}\right)+\ln(\ell+1)\right]\frac{2^{\ell}}{\beta^{\ell+1}}\frac{\Gamma(\ell+1)}{\ell+1}.
\end{equation}
The second term inside the square brackets comes from the asymptotic expansion $\Psi(x)\sim\ln(x)$ of the di-gamma function for large argument $x$ \cite{BookKummer}. Combining the result (\ref{DerivativeL1ellInfinite}) with the expression (\ref{M1ellInfinite}) of $\mathcal{M}_{1,\ell\gg1}$, we deduce that,
\begin{equation}\label{L1ellInfinite}
\mathcal{L}_{1,\ell\gg1}=\frac{1}{2}\left(\frac{\rm d}{{\rm d}s}\mathcal{D}_{1,\ell+s}\right)_{s=0,\ell\gg1}-\mathcal{M}_{1,\ell\gg1}\ln\rho_*\approx\frac{1}{2}\ln\left(\frac{2\ell}{\beta\rho_*^2}\right)\frac{2^{\ell}}{\beta^{\ell+1}}\frac{\Gamma(\ell+1)}{\ell+1}.
\end{equation}

\end{document}